\documentclass[10pt,letterpaper,conference]{IEEEtran}
\IEEEoverridecommandlockouts
\usepackage{cite}
\usepackage{amsmath,amssymb,amsfonts}
\usepackage{algorithmic}
\usepackage{graphicx}
\usepackage{textcomp}
\usepackage[usenames,dvipsnames]{xcolor}
\def\BibTeX{{\rm B\kern-.05em{\sc i\kern-.025em b}\kern-.08em
    T\kern-.1667em\lower.7ex\hbox{E}\kern-.125emX}}

\newcommand\complain[3]{}

\newcommand{\icode}[1]{{\lstset{basicstyle=\ttfamily\small}\lstinline@#1@}}
\newcommand{\icodefns}[1]{{\lstset{basicstyle=\ttfamily\footnotesize}\lstinline@#1@}}

\newcommand{\ie}{\textit{i.e.{}}}
\newcommand{\eg}{\textit{e.g.{}}}
\newcommand{\etal}{\textit{et~al.{}}}

\newcommand{\cival}[2]{\textcolor{gray}{\relsize{-2}{$\pm$#1}}}

\newcommand{\Parallel}{Parallel\xspace}
\newcommand{\GOne}{G1\xspace}
\newcommand{\Epsilon}{Epsilon\xspace}
\newcommand{\Serial}{Serial\xspace}
\newcommand{\ZGC}{ZGC\xspace}
\newcommand{\Shenandoah}{Shenandoah\xspace}

\newcommand{\mdc}{minimum distilled cost\xspace}
\newcommand{\hfac}[1]{${#1}\times$}

\usepackage{listings}

\usepackage{microtype}
\usepackage{balance}
\usepackage{xspace}
\usepackage{mathptm}

\usepackage{array}
\usepackage{caption}
\usepackage{subcaption}
\usepackage{booktabs}
\usepackage{footnote}
\usepackage{multirow}
\usepackage{tablefootnote}
\makeatletter\let\expandableinput\@input\makeatother
\usepackage{relsize}

\newtheorem{definition}{Definition}[section]

\usepackage[binary-units]{siunitx}

\usepackage[inline,shortlabels]{enumitem}

\usepackage[bookmarksnumbered,unicode,hidelinks]{hyperref}
\usepackage[nameinlink,capitalise]{cleveref}

\newcolumntype{B}{>{\global\let\currentrowstyle\relax}} %
\newcolumntype{^}{>{\currentrowstyle}}
\newcommand{\rowstyle}[1]{\gdef\currentrowstyle{#1}#1\ignorespaces}
\newcolumntype{C}{>{\bfseries}}

\usepackage{comment}
\usepackage[normalem]{ulem}

\begin{document}

\title{Distilling the Real Cost of \\ Production Garbage Collectors
\thanks{We thank the anonymous reviewers and Ben Titzer for their detailed feedback and insightful suggestions for improving the paper.
This material is based upon work supported by the Australian Research Council under Grant No.~DP190103367
and National Science Foundation under Grant No.~CSR-2106117.
Zixian Cai is supported by an Australian Government Research Training Program Scholarship.}
}

\author{
    \IEEEauthorblockN{Zixian Cai}
    \IEEEauthorblockA{Australian National University\\
    \href{mailto:zixian.cai@anu.edu.au}{zixian.cai@anu.edu.au}
    }
    \and
    \IEEEauthorblockN{Stephen M.\ Blackburn}
    \IEEEauthorblockA{Australian National University, Google\\
    \href{mailto:steveblackburn@google.com}{steveblackburn@google.com}
    }
    \and
    \IEEEauthorblockN{Michael D.\ Bond}
    \IEEEauthorblockA{Ohio State University\\
    \href{mailto:mikebond@cse.ohio-state.edu}{mikebond@cse.ohio-state.edu}
    }
    \and
    \IEEEauthorblockN{Martin Maas}
    \IEEEauthorblockA{Google\\
    \href{mailto:mmaas@google.com}{mmaas@google.com}
    }
}

\maketitle
\thispagestyle{plain}
\pagestyle{plain}

\begin{abstract}
Despite the long history of garbage collection~(GC) and its prevalence in modern programming languages, there is surprisingly little clarity about its true cost.
Without understanding their cost, crucial tradeoffs made by garbage collectors (GCs) go unnoticed.
This can lead to misguided design constraints and evaluation criteria used by GC researchers and users, hindering the development of high-performance, low-cost GCs.

In this paper, we develop a methodology that allows us to empirically estimate the cost of GC for any given set of metrics.
This fundamental quantification has eluded the research community, even when using modern, well-established methodologies.
By distilling out the explicitly identifiable GC cost, we estimate the intrinsic application execution cost using different GCs.
The \mdc forms a baseline.
Subtracting this baseline from the total execution costs, we can then place an empirical lower bound on the absolute costs of different GCs.
Using this methodology, we study five production GCs in OpenJDK~17, a high-performance Java runtime.
We measure the cost of these collectors, and expose their respective key performance tradeoffs.

We find that with a modestly sized heap,
production GCs incur substantial overheads across a diverse suite of modern benchmarks, spending \emph{at least} 7--82\,\% more wall-clock time and 6--92\,\% more CPU cycles relative to the baseline cost.
We show that these costs can be masked by concurrency and generous provisioning of memory/compute.
In addition, we find that newer low-pause GCs are significantly more expensive than older GCs, and, surprisingly, sometimes deliver \emph{worse} application latency than stop-the-world GCs.

Our findings reaffirm that GC is by no means a solved problem and that a low-cost, low-latency GC remains elusive.
We recommend adopting the distillation methodology together with a wider range of cost metrics for future GC evaluations.
This will not only help the community more comprehensively understand the performance characteristics of different GCs, but also reveal opportunities for future GC optimizations.
\end{abstract}


\section{Introduction}
\label{sec:intro}
Garbage collection~(GC) is ubiquitous in software systems. 
Managed languages, such as C\#, Java, and JavaScript, continue to grow in popularity due to their productivity and safety benefits, which are in part provided by GC.
On servers, many widely used web services, such as Twitter, GitHub, Shopify, and Alibaba, make extensive use of such languages.
On clients, JavaScript engines are embedded in every web browser, and Java runtimes are embedded in every Android phone.

Because of the ubiquity of GC, the research community has extensively studied GC performance.
The approaches include characterizing specific elements of GC behavior, performing comparative evaluation among garbage collectors (GCs), and deconstructing the performance of specific GCs.
These aspects are addressed by a substantial literature, including~\cite{methodology-cacm-2008,mmtk-sigmetrics-2004,barrier-zorn-1990,barrier-oopsla-1992,barrier-ismm-2004,barrier-ismm-2012,locality-oopsla-2004,conservative-spe-1993,dacapo-oopsla-2006,g1-vee-2020}.
They are explicitly non-goals of our work.

While this rich literature helps us understand how GCs compare, how they are designed, and what key sources of cost are, there is a surprising lack of clarity regarding the \emph{real} costs that GC brings to a programming language.
In this paper, we focus on two key problems:
\begin{enumerate*}[(a)]
    \item lack of clarity about the absolute cost of GC, and
    \item misinterpretations of GC evaluations due to limited cost metrics.
\end{enumerate*}
We now offer more detail regarding these two problems and outline our contributions.

\paragraph{Unclear absolute costs}
The absolute cost that garbage collectors impose on modern production runtimes is an important quantification, but it has eluded the community to date.
Its importance is twofold.
For programming language implementers and hardware architects, understanding the absolute cost of GC and its magnitude relative to the rest of the language runtime can help them decide where to spend research and engineering resources.
For language users, knowing the absolute cost of GC can help them decide whether to use a managed language or to use alternatives such as C/C++ and Rust---a decision that often cannot be easily reversed.

Hertz and Berger~\cite{gcmalloc-oopsla-2005} attempted to answer this question, but their work is limited by the fidelity of the simulation infrastructure, and by requiring invasive changes to the runtime.
Due to this complexity, their method cannot be readily applied to modern workloads and production runtimes, and their particular analysis is now somewhat dated by advances in language implementation and computer architecture, and substantial changes to workloads.

\begin{figure*}[t!bp]
    \centering
    \includegraphics[width=\textwidth]{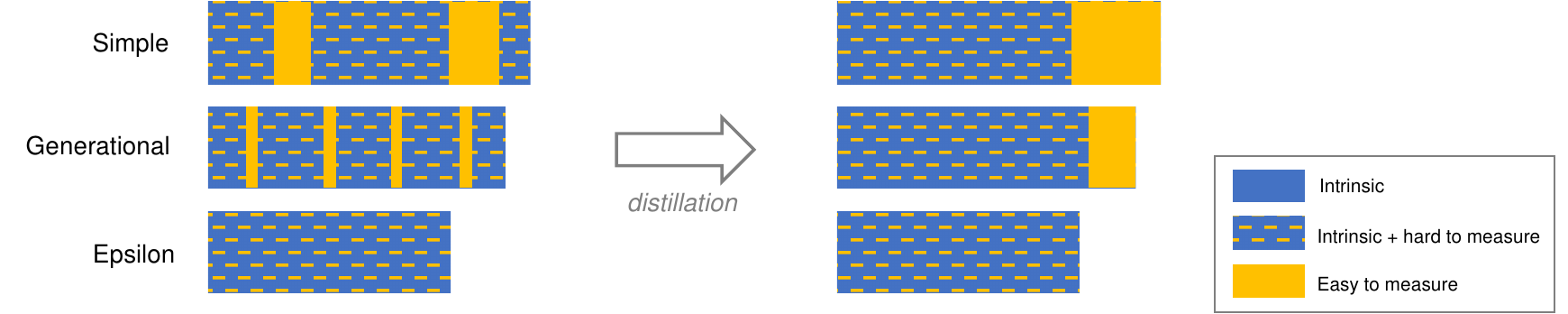}
    \caption{Application execution is comprised of the intrinsic application cost shown in blue and GC costs shown in orange.
    Some of the GC costs, shown with dashed lines, tightly couple with the execution and are hard to measure.
    The distillation process subtracts the easy-to-measure GC costs from the total cost, and what remains approximates the intrinsic application cost.
    The minimum distilled cost can be used to estimate the absolute cost of each GC.}
    \label{fig:distill}
\end{figure*}
In this paper, we develop a language- and runtime-agnostic methodology to empirically place a lower bound on the absolute cost of GC for any cost metric.
The intuition behind our methodology is simple.
If we knew the intrinsic cost of running an application without any of the costs of a garbage collector, then we could use that as a baseline for understanding the absolute cost imposed by real garbage collectors.
However, in practice, some of the GC costs permeate the execution process, and are hard to tease out~(\cref{fig:distill}).
The key insight is that we can approximate this baseline by running an application with real collectors, and distilling out explicitly identifiable GC costs from the total execution costs.
The \emph{minimum distilled cost} overestimates the baseline, and can then be used to derive an empirical lower bound on the absolute cost of each GC.

We analyze all five native GCs in OpenJDK~17, the latest release of an industrial-strength, high-performance JVM.
Our methodology is simple yet effective: even an underestimate of the GCs' absolute costs reveals that they impose substantial costs to program execution across a diverse set of workloads.
We used a wide range of cost metrics including CPU cycles, cache misses, and RAPL energy.
Due to space constraints, we focus on two important metrics, the wall-clock time and CPU cycles, in our case study.
Using a modest heap size, our methodology estimates that by using GC, applications spend 7--82\,\% more wall-clock time and 6--92\,\% more CPU cycles relative to their intrinsic costs.

\paragraph{Misinterpretation of evaluation results}
The second problem we address is that 
GC evaluations are prone to misinterpretation, even when modern, well-established methodologies~(\eg{}, varying the heap size) are used.
Such misinterpretation is dangerous: a costly GC technique understood as cheap can mislead the community to overuse it and dismiss cheaper alternatives, while discouraging future optimizations.
Unless a cost is properly measured, it will not be properly addressed.

This danger is acutely clear in the new low-pause collectors.
These algorithms are popularly understood~\cite{shenandoah-pppj-2016,zgc-slides-2018} to make applications more responsive at low cost 
to throughput.
Our case study~(\cref{sec:misinterpretation}) shows this to be unfounded.
These algorithms impose substantial costs on application execution and do not achieve good application responsiveness. Although they  deliver low \emph{pause times}, the application overheads are so great that responsiveness is often \emph{no better than simple stop-the-world GCs}.

In this paper, we analyze three important types of misinterpretation of GC evaluations arising from overly focusing on limited metrics.
We show how key performance tradeoffs made by different GCs can go unnoticed.
We recommend mitigating this problem by using more cost and performance metrics, including wall-clock time, CPU cycles, and application latency.

\vspace{1.5ex}
\noindent
To summarize, our novel methodology allows the community to grasp the substantial costs incurred by widely used production GCs in real applications.
We recommend empirically estimating the absolute cost of GC, and using richer sets of metrics when evaluating GCs to mitigate common misinterpretations.
These steps are important for revealing the tradeoffs and limitations of different collectors, and for assisting language users in choosing the most appropriate GC for a plethora of existing and emerging workloads.
Our findings invite future research on developing low-latency, low-cost collectors.

The code we used in this work is publicly available: \url{https://github.com/caizixian/distillation}~\cite{cai_zixian_2022_6476821}.

\section{Background and Related Work}
\label{sec:background}
This section discusses a few attempts to measure the absolute cost of GC,
and points out their limitations.
We then highlight two categories of GC costs that are hard to measure: costs tightly coupled with application execution, and indirect costs.
We describe the established methodologies to measure such costs, but they are insufficient to adequately characterize the absolute GC cost on their own.
Finally, we give an overview of the production garbage collectors that we study in this paper, and how they present such hard-to-measure sources of GC cost.

\subsection{Absolute Costs of Garbage Collection}
Hertz~and~Berger~\cite{gcmalloc-oopsla-2005} attempted to quantify the cost of garbage collection over an explicit memory management regime.
Their work shares the same basic intuition as ours~(comparing against a hypothetical ``zero-cost'' baseline), but is an entirely different approach.

By tracing the program execution, they construct a database of object liveness information.
This database allows them to create an ``oracular GC'' that inserts \icode{malloc} and \icode{free} calls into the program execution trace,
which is processed by an architectural simulator (SimpleScalar~\cite{simplescalar-sigarch-1997}).
To do so, they need to make invasive changes to the runtime as well as integrate into SimpleScalar specific information about the behavior of the runtime's collector and allocator.

Although it is an interesting study, it is limited by the fidelity of the simulation infrastructure (a problem whenever using architectural simulation) and by the requirement of being able to make such changes to the runtime and the appropriate modeling within the simulator.
The particular analysis has other limitations including that it is now dated by changes in architecture, workloads and algorithms.

In contrast, our methodology is very simple.
It measures any production collector running on native hardware, making minimal assumptions about the GC implementation, and without access to source code, entirely avoiding the issues in prior work~\cite{gcmalloc-oopsla-2005}.

Some work~(e.g.,~\cite{conservative-spe-1993}) measures the cost of conservative GC in the context of explicit memory management.
In this paper, we focus on precise GC in managed languages.

Numerous studies~(e.g.,~\cite{mmtk-sigmetrics-2004}) measure GC pauses to quantify GC costs.
We show that this approach is problematic because some GC activities~(such as barriers) are carried out by mutators on behalf of the GC, and are hard to separate.
Moreover, for concurrent GCs, while the GC pauses are shorter, the majority of the work is shifted to run concurrently with the mutator threads.

\subsection{Costs Tightly Coupled with Application Execution}
A GC can be considered in terms of three principal activities: \emph{allocation}, \emph{identification}, and \emph{reclamation}~\cite{immix-pldi-2008}.
Allocation is performed by the mutator using space provided by the GC.
Identification establishes which part of the heap is live and which may be reclaimed.
Reclamation makes space occupied by unreachable objects available for reuse.

While some GC activities are performed directly by the collector, others are performed by the mutator; for example:
\begin{enumerate}
    \item Mutators can allocate on their respective local allocation buffers, and only require periodic synchronization with the GC.
    \item Mutators can assist identification and reclamation via \emph{barriers}---code snippets that mediate mutators' heap operations on behalf of the GC~\cite{barrier-ismm-2012}.
    \item Mutators can reclaim unreachable objects.
    For example, in the case of na\"{i}ve reference counting~(RC)~\cite{rc-cacm-1960}, the code that decrements the reference count of an object to zero can immediately reclaim the memory.
\end{enumerate}
For performance reasons, the above mechanisms, such as allocation sequences and barriers, are tightly coupled with mutator execution, often implemented using a ``fast-and-slow-path'' paradigm~\cite{mmtk-icse-2004}.
Fast paths are short but frequently executed code snippets, often inlined into the mutator code by the just-in-time (JIT) compiler.
Because they are so tightly integrated into the mutator, measuring their costs requires carefully crafted methodologies that avoid instrumentation overheads that will affect the measurement:
\begin{enumerate}
    \item Blackburn~\etal{}~\cite{mmtk-sigmetrics-2004} placed an upper bound on the cost of a bump-pointer allocator by compiling the allocation fast path out-of-line.
    Then, that upper bound was used to derive the cost of a free-list allocator using their relative performance.
    \item The cost of barriers has also been extensively studied~\cite{barrier-zorn-1990,barrier-oopsla-1992,barrier-ismm-2004,barrier-ismm-2012}.
    These methodologies remove the requirement of barriers for correctness, \eg{}, through a full-heap trace for a generational GC, and then measure the execution with and without barriers.
    \item Blackburn~\etal{}~\cite{mmtk-sigmetrics-2004} measured the cost of a deferred reference counting collector in terms of the mutator time.
\end{enumerate}
These specialized methodologies require deep understanding of the GC implementation and modifications to the collector and/or runtime, and consequently are not trivially applied to arbitrary language runtimes.
Furthermore, they are insufficient to adequately characterize the absolute GC cost on their own.
In this paper, we focus on devising a methodology that encompasses all of these costs instead of measuring individual components.

\begin{table*}[t!bp]
    \centering
    \caption{Garbage collectors that this paper studies.
    SATB = snapshot at the beginning~\cite{Yuasa90}.
    LVB = loaded value barrier~\cite{TeneIW11}.}
    \sffamily
\begin{tabular}{Br ^l ^l ^l ^l ^l}
\toprule
\rowstyle{\bfseries}
Collector & Year & Generational & Parallel GC & Concurrent GC & Barriers\\
\midrule
Epsilon & 2018 & No~(no GC) & No~(no GC) & No~(no GC) & No~(no GC)\\
Serial & 1998 &Yes & No & No & Write~(card-marking)\\
Parallel & 2005 &Yes & Yes & No & Write~(card-marking) \\
G1~\cite{g1-ismm-2004} & 2009 & Yes & Yes & Yes~(tracing) & Write~(card-marking and SATB)\\
Shenandoah~\cite{shenandoah-pppj-2016} & 2019 & No\footnotemark & Yes & Yes~(tracing and copying) & Write~(SATB) and read~(LVB\footnotemark)\\
ZGC~\cite{zgc-web-2018,zgc-slides-2018} & 2018 & No\footnotemark & Yes & Yes~(tracing and copying) & Read~(LVB)\\
\bottomrule
\end{tabular}

    \label{table:gc}
\end{table*}
\subsection{Indirect Costs and Benefits}

Apart from GC's direct costs, it can also bring indirect costs and even benefits.
A common source of indirect cost (or benefit) is reduced (increased) mutator locality.

GC impacts locality through sharing caches with the mutator.
In the case of concurrent GC, GC threads contend with mutator threads when they share caches.
In the case of stop-the-world pauses, GC code displaces the cache, leaving mutator threads resuming with a cold cache.

However, GC can improve mutator locality through rearranging objects.
A compacting or evacuating GC can improve spatial locality by moving objects that are frequently accessed together to be spatially closer to each other~\cite{locality-oopsla-2004}.

Some of the locality impact might be observable through hardware performance counters, such as cache miss events~(\eg{}, LLC and TLB misses).
However, in general, it is hard to tease out the locality impact of GC running on real hardware, because that would require achieving the same GC effects~(like compacting the heap) without affecting cache state in the process.
It may be possible to achieve such an effect under simulation, but it would likely require significant compute resources to evaluate and would suffer from the same limitations as prior work~\cite{gcmalloc-oopsla-2005}.

In this paper, we are not concerned with teasing out the indirect impact of GC.
Our distillation methodology measures the absolute cost of GC, including its indirect cost.

\subsection{Garbage Collection Algorithms in OpenJDK}
\label{sec:hotspot_gc}

In our case study, we measure the cost of production GCs in OpenJDK~17, and reveal common misinterpretations in their evaluations.
We highlight their key properties in \cref{table:gc}.
Note that analyzing the performance of a particular collector and proposing improvements to its design are explicitly not our objectives.
There exists a substantial literature addressing those, including a recent paper by Zhao and Blackburn that gives a detailed performance analysis of a number of collectors, including ones we study in this paper~\cite{g1-vee-2020}.

The first of these collectors, Epsilon, does not actually collect garbage.
Just like any other collector, its absolute cost---which we find to be nonzero---can be measured by our methodology.
The other five collectors can be divided into three groups:
\begin{enumerate}
    \item Stop-the-world collectors~(\Serial{} and \Parallel{}): the collector requires all mutators to be stopped while it is running, \ie{}, it does not exhibit any concurrency with respect to the mutator.
    \item Concurrent tracing collector~(\GOne{}): the collector performs garbage identification concurrently via a trace.
    This trace does not modify the heap, but marks reachable objects as live.
    The correctness of the concurrent trace is typically protected by write barriers.
    \item Concurrent copying collectors~(\Shenandoah{} and \ZGC{}): in addition to concurrent tracing, the collector performs reclamation concurrently by copying objects.
    This involves modifying the heap and ensuring that the concurrently executing mutator maintains a coherent view of the heap even when objects it references are moved.
    The correctness of concurrent copying is typically protected by read barriers.
\end{enumerate}
All collectors~(except Epsilon) we studied use read and/or write barriers to allow a part of the heap~(a generation or a region) to be independently collected.
Since the main design goals of the two concurrent copying collectors are to reduce the GC pause times and improve the responsiveness for latency-sensitive applications, we also refer to them as the \emph{low-pause collectors} in the rest of the paper.
\footnotetext[1]{In development, see \url{https://openjdk.java.net/jeps/404}.}
\footnotetext[2]{Brooks prior to JDK 13, Baker prior to JDK 14.}
\footnotetext[3]{In development, see \url{https://github.com/openjdk/zgc/tree/zgc_generational}.}

These collectors all include hard-to-measure sources of GC cost discussed previously in this section, hindering the quantification of their absolute costs.
For example, the allocation sequence and the barriers are tightly integrated with the application by inline compilation.
Moreover, when collector threads run concurrently with mutators, it is hard to attribute overheads due to resource contention to the originating threads.   These collectors (including \Epsilon{}) therefore motivate our methodology, with which we are able to measure and place a lower bound on their implicit and explicit costs.

\section{Distilling the Absolute Cost of Garbage Collection}
\label{sec:lbo}

Due to the prevalence of GC, its cost is a hot---and sometimes contentious---topic.
The extensive existing literature focuses on \emph{comparative evaluation}, such as shown in \cref{table:lbo_motivation_total}, which helps us understand how GCs compare.
In this section, we address a distinctly different goal: measuring the absolute cost of GCs.
First, we describe our distillation methodology together with an example.
Then, we discuss the applicability, advantages, and limitations of the methodology.

\subsection{Definition with Examples}

\cref{sec:intro} sketched the intuitions underpinning our distillation methodology~(\cref{fig:distill}).
To deepen the understanding, we now use concrete evaluation results from three production GCs---\Serial{}, \Parallel{}, and \Shenandoah{}---running a real-world application, the H2 database.
This also serves as a running example throughout this section.

Note that in the examples, we use one particular cost metric---total CPU cycles used by all JVM threads---and one particular way of excluding explicitly identifiable GC cost---costs incurred during stop-the-word~(STW) pauses.
We emphasize that these choices are orthogonal to the mechanism of our distillation methodology.
Our methodology can operate on other cost metrics and other ways to isolate GC costs.

It is easy to measure the total costs of real collectors and see how they compare.
\Cref{table:lbo_motivation_total} shows that when using \Serial{}, the total cycle usage is the smallest, with \Parallel{} and \Shenandoah{} being 0.2\,\% and 102.3\,\% more expensive, respectively.
This \emph{comparative analysis} shows which GC is the best to use if you are concerned about total CPU cycle consumption, a well-understood methodology in the literature.
However, it does not measure the absolute CPU cycle cost of each \emph{collector alone}.

\begin{table}[htbp]
    \centering
    \caption{The total CPU cycles consumed when running the DaCapo benchmark h2 with a
    \hfac{3} heap~(see \cref{sec:methodology}) using three different collectors.
    Lower is better.
    The cycles are also normalized to the best collector~(Serial) shown in green.}
    \sffamily
\begin{tabular}{l c@{\hspace{2ex}} r c@{\hspace{2ex}} r}
\toprule
\textbf{Collector} && \textbf{Total} &&  \textbf{Normalized}\\
&& \textcolor{gray}{\relsize{-2}{billion cycles}} && \\
\cmidrule{1-1}\cmidrule{3-3}\cmidrule{5-5}
Parallel   &&  $108.33$ && $1.002$\\
Serial     &&  \color{ForestGreen}$108.12$ && \color{ForestGreen}$1.000$\\
Shenandoah &&  $218.72$ && $2.023$\\
\bottomrule
\end{tabular}

    \label{table:lbo_motivation_total}
\end{table}

\begin{definition}[Intrinsic application cost]
We define the intrinsic application cost as the \emph{theoretical ideal} cost of running an application.
This intrinsic cost includes the best GC benefit~(such as improved locality) any configuration could bring~(\ie{}, a collector with tuning parameters) but none of the GC costs~(such as barriers).

For example, if we knew the intrinsic cost of the application H2 (inclusive of the best GC benefit but exclusive of the GC costs) in terms of CPU cycles, we could subtract that from the total cycles used by each real collector, giving us the absolute cost of the respective collector.
\end{definition}

\begin{definition}[Distilled application cost]
Of course, we do not know the intrinsic application cost of a given workload.
Our key insight is that we can approximate the intrinsic application cost by excluding the costs that we can easily ascribe to GC from its total costs.
This insight can perhaps be better understood using the distillation analogy.
The total cost is a mixture of the intrinsic application cost~(after considering the best GC benefit) and the absolute GC cost.
Some of the GC costs can be explicitly identified~(such as the cost incurred during STW pauses), while others are hard to measure~(such as barriers).
We cannot remove all hard-to-measure GC costs from the mixture, but we can easily distill out explicitly identifiable GC costs.
This gives us an overestimate of the intrinsic application cost.
\[
    \text{Distilled cost} \equiv \text{Total cost} - \text{Explicit GC cost}\; \ge\; \text{Intrinsic cost}
\]
\end{definition}

As shown in \cref{table:lbo_motivation_breakdown}, we can easily distill out explicitly identifiable GC cost by excluding the cost during STW pauses, where no mutator activities happen and costs are strictly from GC.
Repeating the distillation process for each of three collectors, we obtain three distilled application costs.

\begin{table}[htbp]
    \centering
    \caption{Distilling out cycles used during stop-the-world~(STW) pauses from the total cycles in \cref{table:lbo_motivation_total}.
    The minimum distilled application cost~(MDC) in cycles is shown in green.
    The MDC value is used to calculate the empirical lower bound  on the cost of each of the collectors in \cref{table:lbo_calculation}.}
    \sffamily
\begin{tabular}{l c@{\hspace{2ex}} r r r}
\toprule
\textbf{Collector} && \textbf{Total} & \textbf{STW} & \textbf{Distilled}\\
          && \multicolumn{3}{c}{\textcolor{gray}{\relsize{-2}{billion cycles}}}\\
\cmidrule{1-1}\cmidrule{3-5}
Parallel   && $108.33$ & $4.46$  & \color{ForestGreen}$108.33-4.46 = 103.87$   \\ %
Serial     && $108.12$ & $2.75$  & $108.12-2.75=105.37$                     \\ %
Shenandoah && $218.72$ & $0.03$  & $218.72-0.03=218.69$                      \\ %
\bottomrule
\end{tabular}

    \label{table:lbo_motivation_breakdown}
\end{table}

\begin{definition}[Minimum distilled application cost]
The \emph{\mdc}~(MDC) is the minimum of the distilled application costs from running the application with each of the collectors.
Since we define the intrinsic application cost to include the best benefit any GC configuration could bring, the set of collectors used to derive the MDC can include the same collector with \emph{different} tuning parameters~(\eg{}, heap size, the number of collector threads).
The MDC is the best overestimate of the intrinsic cost.
\[
    \left( \text{MDC} \equiv \min_{g\in \text{GCs}}\text{Distilled cost}_g \right) \; \ge \; \text{Intrinsic cost}
\]
\end{definition}
In Table~\ref{table:lbo_motivation_breakdown}, the MDC is 103.87 billion cycles (\Parallel{}, shown in green).

\begin{definition}[Lower Bound Overhead of Garbage Collectors]
The MDC in turn allows us to place an empirical lower bound on the absolute cost  of the collectors, which we call the LBO.
Since $\text{Absolute GC cost} = \text{Total cost} - \text{Intrinsic cost}$ and $\text{MDC} \ge \text{Intrinsic cost}$, for each GC $g$, we have
\[
    \left( \text{LBO}_g \equiv \text{Total cost}_g - \text{MDC} \right) \le \text{Absolute GC cost}_g
\]
\end{definition}
As shown in \cref{table:lbo_calculation}, we subtract the MDC from the total cycles of each of the GCs, yielding a lower bound on the absolute cycle cost of each GC.

\begin{definition}[Normalized LBO]
We can normalize the LBO to the MDC:
\[
    \text{Normalized LBO~(NLBO)} \equiv \text{LBO} / \text{MDC}
\]
\end{definition}
In the above example, we can see that in the context of the H2 database running on OpenJDK, we spend at least 4.1\,\%, 4.3\,\%, and 110.6\,\% more CPU cycles than the intrinsic costs for \Serial, \Parallel, and \Shenandoah{} respectively.
We note that even though \Shenandoah{} is not useful when calculating the MDC, the absolute cost of \Shenandoah{} is still accurately captured.
\begin{table}[htbp]
    \centering
    \caption{The empirical lower bound (LBO) on the cycle cost for each collector can be obtained by subtracting the MDC from the respective \textsf{total} cycles in \cref{table:lbo_motivation_breakdown}.
    The lowest LBO in green.}
    \sffamily
\begin{tabular}{l c@{\hspace{2ex}} r c@{\hspace{2ex}} r}
\toprule
\textbf{Collector} && \multicolumn{1}{c}{\textbf{LBO}} && \textbf{Normalized LBO}\\
          && \multicolumn{1}{c}{\textcolor{gray}{\relsize{-2}{$\text{Total} - \text{MDC}$}}} && \multicolumn{1}{c}{\textcolor{gray}{\relsize{-2}{$\text{LBO} / \text{MDC}$}}}\\
\cmidrule{1-1}\cmidrule{3-3}\cmidrule{5-5}
Parallel   &&  $108.33 - 103.87 =4.46$ && $1.043$\\
Serial     &&  \color{ForestGreen}$108.12-103.87 =4.25$ && \color{ForestGreen}$1.041$\\
Shenandoah &&  $218.72-103.87 =114.85$ && $2.106$\\
\bottomrule
\end{tabular}

    \label{table:lbo_calculation}
\end{table}

We emphasize that the distilled cost of any collector can be calculated, and it is \emph{not} the same as the cost of a no-GC scheme~(such as \Epsilon{} in OpenJDK).
In the above example, all three collectors perform collection, and we estimate their respective LBOs without using a no-GC scheme.

\subsection{Discussion}
The biggest advantage of our distillation methodology is its simplicity.
The only requirement of our methodology is to exclude explicitly identifiable GC cost.

This simplicity makes our methodology generally applicable on different runtimes and GC algorithms.
In the above example, we exclude the costs incurred during STW pauses, where the costs are strictly from GC.
This is conceptually easy, and can be simply implemented as callbacks to delineate STW phases.
In fact, for OpenJDK, this instrumentation can be implemented using JVMTI callbacks~\cite{jvmti-web-2021}, and for .NET, implemented on top of GCRealTimeMon~\cite{realmon-web-2021}.

This simplicity also allows us to evaluate GCs running on native hardware.
As a result, any hardware performance counter can trivially be used as a cost metric when using our methodology.
This includes RAPL energy readings, cache misses, and other metrics.

However, language implementers need to be careful when applying our methodology to GCs in different runtimes.
For example, a compiler that performs escape analysis can lower the GC costs despite no change to the GCs.
Such analysis allows objects to be allocated on the stack rather than on the heap when the objects do not escape.
This reduces overall allocation, and therefore reduces the pressure on the GC.

We also note that to get a useful MDC, our methodology requires \emph{at least} one GC where the distilled cost is close to the intrinsic cost.
For example, if all evaluated collectors were concurrent \emph{and} if explicitly identifiable GC cost were from STW pauses, the distilled costs from \emph{all} collectors would include costs from the concurrently running GC threads.
In this scenario, the MDC would be much higher than the intrinsic cost, and would lead to a very poor empirical lower bound on GC cost.

For hardware performance counters, we can address the above deficiency by measuring the costs on a per-thread basis.
Instead of excluding STW costs, one can exclude costs originating from GC threads.\footnote{Excluding costs of GC threads is not the same as excluding all GC costs. For example, the cost of barriers will still be in the distilled cost.}
This measurement mode is easily supported by, \eg{}, Linux's perf event subsystem.

In our case study, we do not use this engineering optimization.
Because we have two STW collectors~(\Serial{}/\Parallel{}) in the set we studied, taking the whole-process reading at STW points is sufficient to factor out the explicitly identifiable GC cost to establish an MDC that closely approximates the intrinsic application cost.

\section{Case Study: Collectors in OpenJDK 17}
\label{sec:evaluation}
Recall that we address two key problems in this paper:
\begin{enumerate*}
    \item lack of clarity about the costs imposed by GC, and
    \item misinterpretation of GC evaluations due to limited cost metrics.
\end{enumerate*}
In the previous section, we proposed the distillation methodology to estimate the cost of GCs for any given set of metrics.
In this section, we perform a case study of GCs in OpenJDK 17~(\cref{sec:hotspot_gc}), the latest release of a high-performance production JVM.

First, we apply our distillation methodology on these GCs, and measure their costs using the lower bound overhead~(LBOs) defined in \cref{sec:lbo}.
We focus on two cost metrics---the wall-clock time~(\emph{time LBOs}) and the total CPU cycles~(\emph{cycle LBOs}).
We reveal that with a modest \hfac{2.4} heap and across workloads, production collectors can incur substantial absolute costs, evident in both high time LBO and high cycle LBOs.
A surprising trend is that in addition to being significantly slower and cycle intensive, the low-pause collectors fail to deliver better application latency for the evaluated workloads.

Second, we concretely show how simplistic evaluations of these GCs can be misinterpreted.
We highlight three important types of misinterpretation:
\begin{enumerate*}
    \item not considering opportunity cost,
    \item not considering overheads due to concurrency,
    and
    \item measuring pause time instead of application latency.
\end{enumerate*}
We then demonstrate how to mitigate the misinterpretation by using a richer set of cost and performance metrics.

Based on our observations, we recommend that evaluations of GCs should
\begin{enumerate*}
    \item use the distillation methodology to report the costs of GCs, and
    \item use more performance and cost metrics in order to evaluate GCs holistically.
\end{enumerate*}

\subsection{Methodology}
\label{sec:methodology}

\paragraph{Benchmarks and latency measures}

We use a snapshot release of the forthcoming Chopin release of the DaCapo benchmark suite~\cite{dacapo_group_2021_6475255}.
We exclude the benchmarks cassandra, h2o, and kafka, because they use deprecated Java features that are not compatible with JDK 17.

DaCapo's Chopin snapshot release includes a number of latency-sensitive benchmarks.
In these benchmarks,
latency-sensitive services handle remotely issued requests (\eg, over a network) that arrive at some remotely determined \emph{rate}.  When such a system is unable to process a request immediately, it is placed in a queue. The latency of a  request is impacted by three major sources of delay: the uninterrupted time taken to compute the request; the time taken inclusive of interruptions such as GC and scheduling; and the time taken inclusive of interruptions and queuing.
For these latency-sensitive benchmarks, DaCapo reports two measures of latency, \emph{simple} and \emph{metered}, 
DaCapo's \emph{simple} latency ignores queuing, while \emph{metered} latency models requests coming at a metered rate with an arbitrary-sized queue.  When an interruption such as a GC pause occurs, the metered measure reflects the delay that this imposes not only on the currently executing requests, but also on those that are enqueued during the delay.   We use metered latency here because it more accurately models latency-sensitive services.

We also investigate the SPECjbb2015 benchmark, which is often used in the literature when evaluating low-pause GCs.
However, unlike the DaCapo suite, SPECjbb2015 fixes the workload to a constant amount of time rather than to constant work, preventing us from easily measuring the costs of different GCs executing the same amount of work.
Therefore, SPECjbb2015 is not included in our analysis.

\paragraph{Cost metrics}

We implement our distillation methodology as a Java Virtual Machine Tool Interface~(JVMTI)~\cite{jvmti-web-2021} agent.
We implement the distilled cost in the methodology by excluding costs incurred in stop-the-world~(STW) pauses.
The STW points are provided by the JVMTI callbacks for the starts and the ends of GC pauses.

In addition to the wall-clock time, our JVMTI agent captures a wide range of hardware performance counters, including CPU cycles, instruction counts, cache misses, and Intel RAPL energy measurements.
To read these counters, we use the \icode{perf\_events} subsystem in the Linux kernel.
In this section, we focus on two important metrics: the wall-clock time and CPU cycles.

\paragraph{JVM parameters}
\begin{table}[htbp]
    \centering
    \caption{Minimum heap size required to run each benchmark.}
    \sffamily
\begin{tabular}{l c r}
\toprule
\textbf{Benchmark}&& \multicolumn{1}{c}{\textbf{Heap size (MB)}}\\
\cmidrule{1-1}\cmidrule{3-3}
avrora     & & 7   \\
batik      & & 189 \\
biojava    & & 95  \\
eclipse    & & 411 \\
fop        & & 15  \\
graphchi   & & 255 \\
h2         & & 773 \\
jme        & & 29  \\
jython     & & 25  \\
luindex    & & 42  \\
lusearch   & & 21  \\
pmd        & & 156 \\
sunflow    & & 29  \\
tomcat     & & 21  \\
tradebeans & & 131 \\
tradesoap  & & 103 \\
xalan      & & 8   \\
zxing      & & 97  \\
\bottomrule
\end{tabular}

    \label{table:bm_stats}
\end{table}
In this paper, we focus on the out-of-the-box performance characteristics of GCs, and \emph{deliberately do not} set any GC-related parameters except for the heap size.
GC tuning is often specific to particular (classes of) workloads, whereas our concern is the real cost of each GC when handling a diverse set of workloads.
Also, in general, GC tuning is an open-ended problem, outside the scope of this paper.
We report the performance of each GC for different heap sizes, because the performance of GC is sensitive to the heap size~\cite{dacapo-oopsla-2006}.
Except for Epsilon, which does not perform GC, we set the heap size relative to the minimum heap size~(\cref{table:bm_stats}) required to run each benchmark~(\eg, \hfac{2.0} means that the heap size is set to be twice as big as the minimum heap required for a particular benchmark).
The minimum heap size for each benchmark is measured using \GOne because it is the most space-efficient GC among the ones we study.

The only other JVM parameters we set are \icode{-server -XX:-TieredCompilation -Xcomp},
to speed up the warmup of the JVM and reduce the experimental noise due to JIT compilation.
We omit parameters \icode{-XX:-TieredCompilation -Xcomp} for tradebeans and tradesoap, because these parameters cause these two benchmarks to crash for the version of OpenJDK we use.

\paragraph{Execution methodology}
For each configuration, we invoke each benchmark for 20 times.  We interleave invocations of different configurations to minimize bias due to systemic interference.
In each \emph{invocation}, the benchmark performs five \emph{iterations}, and we report results from the last iteration, with the first four iterations serving to warm up the runtime.
For each configuration, we report the mean and the 95\,\% confidence interval~(CI) based on the 20 invocations.

\paragraph{Hardware}
While microarchitectural sensitivity of a GC algorithm is important, it is orthogonal to this work.
We use two identical machines with Intel Core i9-9900K~(Coffee Lake) CPUs~(8 cores, 16 threads), with 4$\times$32G DDR4-3200 memory.
We disable the dynamic frequency scaling~(\ie, Turbo Boost) to reduce  experimental noise.

\paragraph{Software}
All machines run identical Ubuntu 18.04.6 LTS images with \icode{5.4.0-89-generic} kernels.
We use the \icode{Temurin-17.0.1+12} release of OpenJDK,
which is Eclipse Temurin's (formerly AdoptOpenJDK's) distribution of OpenJDK.
We focus on OpenJDK 17 in this paper, as it is the latest LTS release as of writing and is supposed to bring many GC performance improvements~(such as~\cite{zgc17-blog-2021,shenandoah17-blog-2021}) since OpenJDK 11.
We also measured OpenJDK 11 using the \icode{AdoptOpenJDK-11.0.11+9} release, and the overall results were quite similar to those of OpenJDK 17.
All benchmarks are executed on an otherwise idle machine, with as many background daemons and periodic tasks disabled as possible.

\begin{table}[tbp]
    \centering
    \caption{Time LBOs averaged over 16 benchmarks.
    The best value for each heap size is shown in green.
    Where a collector cannot run all benchmarks at a particular heap size, the entry is left blank.
    \Parallel outperforms other collectors, except at \hfac{1.4}, where \GOne has the lowest cost.
    }
    \sffamily
\setlength{\tabcolsep}{5.2pt} %
\begin{tabular}{l c@{\hspace{0.5em}} ^r ^r ^r ^r ^r ^r ^r ^r}
\toprule
\textbf{GC} && \hfac{1.4} & \hfac{1.9} & \hfac{2.4} & \hfac{3.0} & \hfac{3.7} & \hfac{4.4} & \hfac{5.2} & \hfac{6.0}\\
\cmidrule{1-1}\cmidrule{3-10}
Ser. && 1.42 & 1.17 & 1.14 & 1.13 & 1.11 & 1.10 & 1.09 & 1.09 \\
Par. && 1.41 & \color{ForestGreen}1.09 & \color{ForestGreen}1.07 & \color{ForestGreen}1.06 & \color{ForestGreen}1.05 & \color{ForestGreen}1.04 & \color{ForestGreen}1.04 & \color{ForestGreen}1.03 \\
G1 && \color{ForestGreen}1.24 & 1.16 & 1.11 & 1.09 & 1.08 & 1.07 & 1.07 & 1.06 \\
Shen. && * & 1.94 & 1.64 & 1.43 & 1.37 & 1.30 & 1.25 & 1.23 \\
ZGC && * & * & 1.82 & 1.54 & 1.39 & 1.32 & 1.27 & 1.23 \\
\bottomrule
\end{tabular}

    \label{table:time_lbo}
\bigskip
    \centering
    \caption{Cycle LBOs averaged over 16 benchmarks.
    The best value for each heap size is shown in green.
    Where a collector cannot run all benchmarks at a particular heap size, the  entry is left blank.
    \Serial consistently outperforms other collectors for all heap sizes.
    }
    \sffamily
\setlength{\tabcolsep}{5.2pt} %
\begin{tabular}{l c@{\hspace{0.5em}} ^r ^r ^r ^r ^r ^r ^r ^r}
\toprule
\textbf{GC} && \hfac{1.4} & \hfac{1.9} & \hfac{2.4} & \hfac{3.0} & \hfac{3.7} & \hfac{4.4} & \hfac{5.2} & \hfac{6.0}\\
\cmidrule{1-1}\cmidrule{3-10}
Ser. && \color{ForestGreen}1.22 & \color{ForestGreen}1.08 & \color{ForestGreen}1.06 & \color{ForestGreen}1.06 & \color{ForestGreen}1.04 & \color{ForestGreen}1.04 & \color{ForestGreen}1.04 & \color{ForestGreen}1.04 \\
Par. && 1.70 & 1.15 & 1.12 & 1.10 & 1.07 & 1.06 & 1.06 & 1.05 \\
G1 && 1.54 & 1.34 & 1.17 & 1.14 & 1.10 & 1.09 & 1.09 & 1.09 \\
Shen. && * & 1.75 & 1.54 & 1.47 & 1.42 & 1.39 & 1.36 & 1.33 \\
ZGC && * & * & 1.92 & 1.68 & 1.55 & 1.46 & 1.40 & 1.34 \\
\bottomrule
\end{tabular}

    \label{table:cycles_lbo}
\end{table}

\subsection{Results: Costs of Garbage Collection}
\label{sec:gc_overhead}

First, we estimate the costs of each \emph{configuration}~(a collector at a heap size) using our distillation methodology.
The distilled cost for each configuration is the total cost excluding the STW cost.
Note that Epsilon does not actually collect garbage, and therefore, its distilled cost is the same as the total cost.

Though \Epsilon is not a practical GC, its distilled cost could potentially help us obtain an MDC that better estimates the intrinsic cost of a workload.
However, in practice, \Epsilon has high absolute costs because the spatial locality of objects is poor, and the allocation sequence will always need to request more physical memory from the operating system.
This is evident in Tables~\ref{table:time_lbo_3023} and \ref{table:cycles_lbo_3023}, showing the LBOs of \Epsilon beside those of standard collectors, for cases in which \Epsilon is able to run a benchmark without exhausting the machine's available memory.
From the tables, we can see that \Epsilon only affects the MDC of one benchmark---sunflow---indicated by the $1.000$ time LBO and the $1.000$ cycle LBO.
For the rest of the section, we focus on the absolute costs of collectors other than Epsilon.

\Cref{table:time_lbo} shows the time LBOs of the collectors we study while \cref{table:cycles_lbo} shows the cycle LBOs.
Each table covers eight different heap sizes, ranging from a small \hfac{1.4} heap to a generous \hfac{6.0} heap~(see \cref{sec:methodology} for the multiplier notation).
The LBO value for each configuration is the geometric mean over 16
DaCapo benchmarks.\footnote{We use 18 DaCapo benchmarks in total.
However, eclipse and xalan are excluded in the geometric mean calculation because too many collectors were not able to run these two benchmarks for small heap sizes.
If we were to include these two benchmarks, a lot more entries would be missing from the tables.}

The production GCs incur substantial costs.
For a modest \hfac{2.4} heap, production GCs on average spend at least 7--82\,\% more wall-clock time and 6--92\,\% more cycles relative to the intrinsic costs.
Even for a generous \hfac{6.0} heap, the costs are as much as 23\,\% in terms of time and 34\,\% in terms of cycles.
For all heap sizes shown, \Serial achieves the lowest costs in terms of cycles, while \Parallel achieves the lowest costs in terms of time for all but the smallest heap size.

Note that even though \Serial is single-threaded, the rest of the VM is still multithreaded, hence the difference in between the time and cycle LBOs for \Serial.
Unsurprisingly, this difference is more pronounced on benchmarks with more mutator parallelism.

Tables~\ref{table:time_lbo_3023} and \ref{table:cycles_lbo_3023} take a closer look at the \hfac{3.0} heap size.
This allows us to observe how different collectors behave when challenged with a diverse, modern workload.
\begin{table}[tbp]
    \centering
    \caption{Time LBOs at \hfac{3} heap~(except \Epsilon) using distillation.
    Lower is better.
    xalan is excluded from the summary statistics due to ZGC failing to run, and the corresponding row is grayed out.
    The best results for each benchmark are shown in green~(light green for xalan).
    Each LBO is the mean of 20 invocations, with its 95\,\% CI shown below in gray.
    \Parallel outperforms other collectors for most benchmarks.}
    \sffamily
\begin{tabular}{l c@{\hspace{0.2ex}} r r c@{\hspace{0.2ex}} r  c@{\hspace{0.2ex}} r r c@{\hspace{0.2ex}} r}
\toprule
\textbf{Bench.} && \textbf{\Serial{}} & \textbf{Para.} && \textbf{\GOne{}} && \textbf{Shen.} & \textbf{\ZGC{}} && \textbf{Eps.}\\
\cmidrule{1-1}\cmidrule{3-4}\cmidrule{6-6}\cmidrule{8-9}\cmidrule{11-11}
avrora &  & \color{ForestGreen}1.009 & 1.025 &  & 1.031 &  & 1.101 & 1.072 &  & 1.010\\
 &  & \cival{0.7\,\%}{0.7\,\%} & \cival{0.5\,\%}{0.5\,\%} &  & \cival{0.6\,\%}{0.6\,\%} &  & \cival{0.8\,\%}{0.8\,\%} & \cival{0.7\,\%}{0.7\,\%} &  & \cival{0.7\,\%}{0.7\,\%}\\
batik &  & 1.146 & 1.092 &  & \color{ForestGreen}1.067 &  & 1.089 & 1.077 &  & 1.087\\
 &  & \cival{0.2\,\%}{0.2\,\%} & \cival{0.4\,\%}{0.4\,\%} &  & \cival{1.0\,\%}{1.0\,\%} &  & \cival{0.4\,\%}{0.4\,\%} & \cival{0.3\,\%}{0.3\,\%} &  & \cival{0.2\,\%}{0.2\,\%}\\
biojava &  & \color{ForestGreen}1.006 & 1.007 &  & 1.033 &  & 1.221 & 1.882 &  & 1.157\\
 &  & \cival{0.1\,\%}{0.1\,\%} & \cival{0.1\,\%}{0.1\,\%} &  & \cival{0.2\,\%}{0.2\,\%} &  & \cival{0.2\,\%}{0.2\,\%} & \cival{1.0\,\%}{1.0\,\%} &  & \cival{0.1\,\%}{0.1\,\%}\\
eclipse &  & 1.050 & \color{ForestGreen}1.010 &  & 1.062 &  & 1.127 & 1.136 &  & 1.180\\
 &  & \cival{0.4\,\%}{0.4\,\%} & \cival{0.1\,\%}{0.1\,\%} &  & \cival{0.2\,\%}{0.2\,\%} &  & \cival{0.2\,\%}{0.2\,\%} & \cival{0.1\,\%}{0.1\,\%} &  & \cival{0.3\,\%}{0.3\,\%}\\
fop &  & 1.159 & \color{ForestGreen}1.129 &  & 1.201 &  & 1.414 & 2.107 &  & 1.175\\
 &  & \cival{0.2\,\%}{0.2\,\%} & \cival{1.6\,\%}{1.6\,\%} &  & \cival{1.2\,\%}{1.2\,\%} &  & \cival{0.8\,\%}{0.8\,\%} & \cival{0.6\,\%}{0.6\,\%} &  & \cival{0.3\,\%}{0.3\,\%}\\
graphchi &  & 1.021 & \color{ForestGreen}1.011 &  & 1.020 &  & 1.191 & 1.125 &  & 1.054\\
 &  & \cival{0.3\,\%}{0.3\,\%} & \cival{0.3\,\%}{0.3\,\%} &  & \cival{0.4\,\%}{0.4\,\%} &  & \cival{1.5\,\%}{1.5\,\%} & \cival{0.4\,\%}{0.4\,\%} &  & \cival{0.7\,\%}{0.7\,\%}\\
h2 &  & 1.121 & \color{ForestGreen}1.044 &  & 1.093 &  & 1.570 & 1.964 &  & 1.483\\
 &  & \cival{0.3\,\%}{0.3\,\%} & \cival{0.3\,\%}{0.3\,\%} &  & \cival{0.6\,\%}{0.6\,\%} &  & \cival{0.3\,\%}{0.3\,\%} & \cival{0.6\,\%}{0.6\,\%} &  & \cival{0.2\,\%}{0.2\,\%}\\
jme &  & 1.003 & 1.003 &  & \color{ForestGreen}1.002 &  & 1.010 & 1.006 &  & 1.006\\
 &  & \cival{0.0\,\%}{0.0\,\%} & \cival{0.0\,\%}{0.0\,\%} &  & \cival{0.0\,\%}{0.0\,\%} &  & \cival{0.0\,\%}{0.0\,\%} & \cival{0.1\,\%}{0.1\,\%} &  & \cival{0.0\,\%}{0.0\,\%}\\
jython &  & 1.037 & \color{ForestGreen}1.022 &  & 1.065 &  & 1.609 & 2.204 &  & 1.231\\
 &  & \cival{1.7\,\%}{1.7\,\%} & \cival{0.2\,\%}{0.2\,\%} &  & \cival{0.2\,\%}{0.2\,\%} &  & \cival{0.9\,\%}{0.9\,\%} & \cival{0.5\,\%}{0.5\,\%} &  & \cival{0.3\,\%}{0.3\,\%}\\
luindex &  & \color{ForestGreen}1.007 & 1.011 &  & 1.044 &  & 1.150 & 1.097 &  & 1.055\\
 &  & \cival{0.2\,\%}{0.2\,\%} & \cival{0.3\,\%}{0.3\,\%} &  & \cival{0.4\,\%}{0.4\,\%} &  & \cival{0.7\,\%}{0.7\,\%} & \cival{0.3\,\%}{0.3\,\%} &  & \cival{0.8\,\%}{0.8\,\%}\\
lusearch &  & 1.230 & \color{ForestGreen}1.105 &  & 1.135 &  & 3.766 & 2.701 &  & \\
 &  & \cival{0.2\,\%}{0.2\,\%} & \cival{0.3\,\%}{0.3\,\%} &  & \cival{0.4\,\%}{0.4\,\%} &  & \cival{1.6\,\%}{1.6\,\%} & \cival{0.4\,\%}{0.4\,\%} &  & \\
pmd &  & 1.613 & \color{ForestGreen}1.124 &  & 1.188 &  & 1.559 & 1.777 &  & 1.165\\
 &  & \cival{0.2\,\%}{0.2\,\%} & \cival{0.2\,\%}{0.2\,\%} &  & \cival{0.3\,\%}{0.3\,\%} &  & \cival{0.6\,\%}{0.6\,\%} & \color{red}\cival{8.7\,\%}{8.7\,\%} &  & \cival{0.4\,\%}{0.4\,\%}\\
sunflow &  & 1.412 & 1.156 &  & \color{ForestGreen}1.156 &  & 2.180 & 2.109 &  & 1.000\\
 &  & \cival{2.0\,\%}{2.0\,\%} & \cival{2.0\,\%}{2.0\,\%} &  & \cival{1.2\,\%}{1.2\,\%} &  & \cival{2.2\,\%}{2.2\,\%} & \cival{0.6\,\%}{0.6\,\%} &  & \cival{0.4\,\%}{0.4\,\%}\\
tomcat &  & 1.157 & \color{ForestGreen}1.137 &  & 1.144 &  & 1.305 & 1.914 &  & 1.145\\
 &  & \cival{0.2\,\%}{0.2\,\%} & \cival{0.2\,\%}{0.2\,\%} &  & \cival{0.1\,\%}{0.1\,\%} &  & \cival{0.2\,\%}{0.2\,\%} & \cival{0.6\,\%}{0.6\,\%} &  & \cival{0.1\,\%}{0.1\,\%}\\
tradebeans &  & 1.120 & \color{ForestGreen}1.020 &  & 1.164 &  & 1.423 & 1.327 &  & 1.415\\
 &  & \cival{0.6\,\%}{0.6\,\%} & \cival{0.6\,\%}{0.6\,\%} &  & \cival{0.7\,\%}{0.7\,\%} &  & \cival{0.8\,\%}{0.8\,\%} & \cival{0.5\,\%}{0.5\,\%} &  & \cival{0.5\,\%}{0.5\,\%}\\
tradesoap &  & 1.094 & \color{ForestGreen}1.011 &  & 1.112 &  & 1.340 & 1.282 &  & 1.376\\
 &  & \cival{0.4\,\%}{0.4\,\%} & \cival{0.2\,\%}{0.2\,\%} &  & \cival{0.3\,\%}{0.3\,\%} &  & \cival{0.2\,\%}{0.2\,\%} & \cival{0.2\,\%}{0.2\,\%} &  & \cival{0.2\,\%}{0.2\,\%}\\
xalan &  & \color{Gray}3.380 & \color{SpringGreen}3.074 &  & \color{Gray}3.496 &  & \color{Gray}30.176 &  &  & \color{Gray}1.111\\
 &  & \cival{0.3\,\%}{0.3\,\%} & \cival{0.4\,\%}{0.4\,\%} &  & \cival{0.3\,\%}{0.3\,\%} &  & \cival{1.7\,\%}{1.7\,\%} &  &  & \cival{0.6\,\%}{0.6\,\%}\\
zxing &  & 1.030 & 1.058 &  & \color{ForestGreen}1.028 &  & 1.274 & 1.235 &  & 1.062\\
 &  & \cival{1.1\,\%}{1.1\,\%} & \cival{1.8\,\%}{1.8\,\%} &  & \cival{1.3\,\%}{1.3\,\%} &  & \cival{1.1\,\%}{1.1\,\%} & \cival{1.1\,\%}{1.1\,\%} &  & \cival{1.3\,\%}{1.3\,\%}\\
\cmidrule{1-1}\cmidrule{3-4}\cmidrule{6-6}\cmidrule{8-9}
\multicolumn{1}{r}{\textbf{min}} &  & \textit{1.003} & \textit{1.003} &  & \textit{1.002} &  & \textit{1.010} & \textit{1.006}\\
\multicolumn{1}{r}{\textbf{max}} &  & \textit{1.613} & \textit{1.156} &  & \textit{1.201} &  & \textit{3.766} & \textit{2.701}\\
\multicolumn{1}{r}{\textbf{mean}} &  & \textit{1.130} & \textit{1.057} &  & \textit{1.091} &  & \textit{1.490} & \textit{1.589}\\
\multicolumn{1}{r}{\textbf{geomean}} &  & \textit{1.121} & \textit{1.055} &  & \textit{1.089} &  & \textit{1.407} & \textit{1.513}\\

\bottomrule
\end{tabular}

    \label{table:time_lbo_3023}
\end{table}

\subsection{Analysis of Results}
\label{sec:gc_overhead_analysis}

We compare the performance trends among and within three groups of GCs~(see \cref{sec:hotspot_gc}).

\paragraph{Stop-the-world~(STW) GCs vs.\ Concurrent GCs}
Overall, STW collectors~(\Serial and \Parallel) are cheaper than concurrent collectors~(\GOne, \Shenandoah, and \ZGC), both in terms of the time and cycles.
The exception is that \Serial is uncompetitive in terms of time due to its  lack of parallelism.
At a \hfac{3.0} heap, in terms of cycles, STW collectors are never more expensive than concurrent collectors.\footnote{For sunflow, \Parallel is 1\,\% more expensive than \ZGC, but the confidence intervals overlap.}
In terms of time, STW collectors are only more expensive than concurrent collectors for 2 out of 17 benchmarks: batik and zxing.\footnote{The differences for jme and sunflow are
negligible (and also not statistically significant for sunflow).}

\paragraph{Single-threaded vs.\ multi-threaded STW GC}
Between two stop-the-world collectors, \Parallel is more expensive than \Serial in terms of cycles~(as much as 48\,\% for a small \hfac{1.4} heap), and the reverse is true in terms of time.
This effect occurs presumably because a multi-threaded collector exploits the available parallelism to shorten pauses but introduces synchronization overhead among collector threads.
At \hfac{3.0} heap, in terms of cycles, \Parallel is only cheaper than \Serial for 2 out of 17 benchmarks: lusearch, tradebeans, and tradesoap.\footnote{The differences for jython and sunflow are not statistically significant.}
In terms of time, \Serial is only cheaper than \Parallel for avrora.\footnote{The differences for biojava, luindex, and sunflow are not statistically significant.}

\begin{table}[htbp]
    \centering
    \caption{Cycle LBOs at \hfac{3} heap~(except \Epsilon) using distillation.
    Lower is better.
    xalan is excluded from the summary statistics due to ZGC failing to run, and the corresponding row is grayed out.
    The best results for each benchmark are shown in green~(light green for xalan).
    Each LBO is the mean of 20 invocations, with its 95\,\% CI shown below in gray.
    \Serial outperforms other collectors for most benchmarks.}
    \sffamily
\begin{tabular}{l c@{\hspace{0.2ex}} r r c@{\hspace{0.2ex}} r  c@{\hspace{0.2ex}} r r c@{\hspace{0.2ex}} r}
\toprule
\textbf{Bench.} && \textbf{\Serial{}} & \textbf{Para.} && \textbf{\GOne{}} && \textbf{Shen.} & \textbf{\ZGC{}} && \textbf{Eps.}\\
\cmidrule{1-1}\cmidrule{3-4}\cmidrule{6-6}\cmidrule{8-9}\cmidrule{11-11}
avrora &  & \color{ForestGreen}1.007 & 1.014 &  & 1.047 &  & 1.201 & 1.209 &  & 1.017\\
 &  & \cival{0.6\,\%}{0.6\,\%} & \cival{0.7\,\%}{0.7\,\%} &  & \cival{0.7\,\%}{0.7\,\%} &  & \cival{0.6\,\%}{0.6\,\%} & \cival{0.7\,\%}{0.7\,\%} &  & \cival{0.4\,\%}{0.4\,\%}\\
batik &  & \color{ForestGreen}1.146 & 1.991 &  & 1.565 &  & 1.454 & 1.915 &  & 1.087\\
 &  & \cival{0.2\,\%}{0.2\,\%} & \cival{2.1\,\%}{2.1\,\%} &  & \cival{0.8\,\%}{0.8\,\%} &  & \cival{0.5\,\%}{0.5\,\%} & \color{red}\cival{3.3\,\%}{3.3\,\%} &  & \cival{0.2\,\%}{0.2\,\%}\\
biojava &  & \color{ForestGreen}1.006 & 1.014 &  & 1.046 &  & 1.521 & 3.962 &  & 1.157\\
 &  & \cival{0.1\,\%}{0.1\,\%} & \cival{0.1\,\%}{0.1\,\%} &  & \cival{0.3\,\%}{0.3\,\%} &  & \cival{0.7\,\%}{0.7\,\%} & \cival{1.0\,\%}{1.0\,\%} &  & \cival{0.1\,\%}{0.1\,\%}\\
eclipse &  & \color{ForestGreen}1.054 & 1.109 &  & 1.317 &  & 1.390 & 1.474 &  & 1.196\\
 &  & \cival{0.4\,\%}{0.4\,\%} & \cival{0.1\,\%}{0.1\,\%} &  & \cival{1.2\,\%}{1.2\,\%} &  & \cival{0.3\,\%}{0.3\,\%} & \cival{0.5\,\%}{0.5\,\%} &  & \cival{0.1\,\%}{0.1\,\%}\\
fop &  & \color{ForestGreen}1.160 & 1.203 &  & 1.460 &  & 1.893 & 2.227 &  & 1.175\\
 &  & \cival{0.3\,\%}{0.3\,\%} & \cival{1.5\,\%}{1.5\,\%} &  & \cival{1.9\,\%}{1.9\,\%} &  & \cival{1.1\,\%}{1.1\,\%} & \cival{1.1\,\%}{1.1\,\%} &  & \cival{0.3\,\%}{0.3\,\%}\\
graphchi &  & \color{ForestGreen}1.008 & 1.031 &  & 1.048 &  & 1.226 & 1.226 &  & 1.099\\
 &  & \cival{0.3\,\%}{0.3\,\%} & \cival{0.3\,\%}{0.3\,\%} &  & \cival{0.3\,\%}{0.3\,\%} &  & \cival{2.2\,\%}{2.2\,\%} & \cival{0.7\,\%}{0.7\,\%} &  & \cival{0.6\,\%}{0.6\,\%}\\
h2 &  & \color{ForestGreen}1.053 & 1.055 &  & 1.123 &  & 2.131 & 2.645 &  & 1.365\\
 &  & \cival{0.1\,\%}{0.1\,\%} & \cival{0.2\,\%}{0.2\,\%} &  & \cival{0.4\,\%}{0.4\,\%} &  & \cival{0.3\,\%}{0.3\,\%} & \cival{0.7\,\%}{0.7\,\%} &  & \cival{0.2\,\%}{0.2\,\%}\\
jme &  & \color{ForestGreen}1.071 & 1.132 &  & 1.091 &  & 1.517 & 1.470 &  & 1.133\\
 &  & \cival{0.3\,\%}{0.3\,\%} & \cival{0.4\,\%}{0.4\,\%} &  & \cival{0.3\,\%}{0.3\,\%} &  & \cival{0.5\,\%}{0.5\,\%} & \cival{1.0\,\%}{1.0\,\%} &  & \cival{0.3\,\%}{0.3\,\%}\\
jython &  & 1.036 & \color{ForestGreen}1.034 &  & 1.072 &  & 2.038 & 2.444 &  & 1.225\\
 &  & \cival{1.7\,\%}{1.7\,\%} & \cival{0.2\,\%}{0.2\,\%} &  & \cival{0.2\,\%}{0.2\,\%} &  & \cival{0.7\,\%}{0.7\,\%} & \cival{0.4\,\%}{0.4\,\%} &  & \cival{0.3\,\%}{0.3\,\%}\\
luindex &  & \color{ForestGreen}1.010 & 1.018 &  & 1.067 &  & 1.207 & 1.199 &  & 1.065\\
 &  & \cival{0.2\,\%}{0.2\,\%} & \cival{0.2\,\%}{0.2\,\%} &  & \cival{1.3\,\%}{1.3\,\%} &  & \cival{0.8\,\%}{0.8\,\%} & \cival{0.5\,\%}{0.5\,\%} &  & \cival{0.9\,\%}{0.9\,\%}\\
lusearch &  & 1.045 & \color{ForestGreen}1.030 &  & 1.081 &  & 1.213 & 1.268 &  & \\
 &  & \cival{0.3\,\%}{0.3\,\%} & \cival{0.2\,\%}{0.2\,\%} &  & \cival{0.2\,\%}{0.2\,\%} &  & \cival{0.3\,\%}{0.3\,\%} & \cival{0.3\,\%}{0.3\,\%} &  & \\
pmd &  & \color{ForestGreen}1.096 & 1.178 &  & 1.286 &  & 1.563 & 1.655 &  & 1.266\\
 &  & \cival{0.5\,\%}{0.5\,\%} & \cival{0.4\,\%}{0.4\,\%} &  & \cival{0.6\,\%}{0.6\,\%} &  & \cival{0.5\,\%}{0.5\,\%} & \color{red}\cival{10.1\,\%}{10.1\,\%} &  & \cival{0.2\,\%}{0.2\,\%}\\
sunflow &  & 1.125 & 1.087 &  & 1.088 &  & 1.240 & \color{ForestGreen}1.076 &  & 1.000\\
 &  & \cival{2.4\,\%}{2.4\,\%} & \cival{2.1\,\%}{2.1\,\%} &  & \cival{1.2\,\%}{1.2\,\%} &  & \cival{0.8\,\%}{0.8\,\%} & \cival{0.8\,\%}{0.8\,\%} &  & \cival{0.4\,\%}{0.4\,\%}\\
tomcat &  & \color{ForestGreen}1.010 & 1.015 &  & 1.022 &  & 1.119 & 1.119 &  & 1.035\\
 &  & \cival{0.1\,\%}{0.1\,\%} & \cival{0.1\,\%}{0.1\,\%} &  & \cival{0.1\,\%}{0.1\,\%} &  & \cival{0.1\,\%}{0.1\,\%} & \cival{0.1\,\%}{0.1\,\%} &  & \cival{0.1\,\%}{0.1\,\%}\\
tradebeans &  & 1.060 & \color{ForestGreen}1.049 &  & 1.211 &  & 1.637 & 2.171 &  & 1.293\\
 &  & \cival{0.4\,\%}{0.4\,\%} & \cival{0.3\,\%}{0.3\,\%} &  & \cival{0.7\,\%}{0.7\,\%} &  & \cival{0.3\,\%}{0.3\,\%} & \cival{0.5\,\%}{0.5\,\%} &  & \cival{0.6\,\%}{0.6\,\%}\\
tradesoap &  & 1.052 & \color{ForestGreen}1.029 &  & 1.117 &  & 1.710 & 2.155 &  & 1.251\\
 &  & \cival{0.4\,\%}{0.4\,\%} & \cival{0.4\,\%}{0.4\,\%} &  & \cival{0.8\,\%}{0.8\,\%} &  & \cival{0.4\,\%}{0.4\,\%} & \cival{0.4\,\%}{0.4\,\%} &  & \cival{0.2\,\%}{0.2\,\%}\\
xalan &  & \color{SpringGreen}1.211 & \color{Gray}1.633 &  & \color{Gray}1.783 &  & \color{Gray}1.744 &  &  & \color{Gray}1.139\\
 &  & \cival{0.4\,\%}{0.4\,\%} & \cival{0.3\,\%}{0.3\,\%} &  & \cival{0.3\,\%}{0.3\,\%} &  & \cival{0.4\,\%}{0.4\,\%} &  &  & \cival{0.3\,\%}{0.3\,\%}\\
zxing &  & \color{ForestGreen}1.017 & 1.034 &  & 1.027 &  & 1.275 & 1.238 &  & 1.061\\
 &  & \cival{1.1\,\%}{1.1\,\%} & \cival{1.8\,\%}{1.8\,\%} &  & \cival{1.4\,\%}{1.4\,\%} &  & \cival{1.2\,\%}{1.2\,\%} & \cival{1.1\,\%}{1.1\,\%} &  & \cival{1.4\,\%}{1.4\,\%}\\
\cmidrule{1-1}\cmidrule{3-4}\cmidrule{6-6}\cmidrule{8-9}
\multicolumn{1}{r}{\textbf{min}} &  & \textit{1.006} & \textit{1.014} &  & \textit{1.022} &  & \textit{1.119} & \textit{1.076}\\
\multicolumn{1}{r}{\textbf{max}} &  & \textit{1.160} & \textit{1.991} &  & \textit{1.565} &  & \textit{2.131} & \textit{3.962}\\
\multicolumn{1}{r}{\textbf{mean}} &  & \textit{1.056} & \textit{1.119} &  & \textit{1.157} &  & \textit{1.490} & \textit{1.791}\\
\multicolumn{1}{r}{\textbf{geomean}} &  & \textit{1.055} & \textit{1.103} &  & \textit{1.148} &  & \textit{1.462} & \textit{1.671}\\

\bottomrule
\end{tabular}

    \label{table:cycles_lbo_3023}
\end{table}
\paragraph{Concurrent tracing vs.\ concurrent copying GC}
Among concurrent collectors, the newer, concurrent copying collectors~(\Shenandoah and \ZGC) are significantly more costly than the concurrent tracing collector~(\GOne).
The difference is up to 75\,\% in terms of cycles (\GOne vs.\ \ZGC at \hfac{2.4} heap) and 78\,\% in terms of time (\GOne vs.\ \Shenandoah at \hfac{1.9} heap).
This is presumably due to the use of costly read barriers and the (un)timeliness of reclamation, which we discuss below.
At \hfac{3.0} heap, \GOne consistently outperforms \Shenandoah and \ZGC for all benchmarks in terms of time.
In terms of cycles, \Shenandoah and \ZGC are only cheaper than \GOne for batik and xalan~(not statistically significant for sunflow).

\paragraph{Pathological Modes of Concurrent Copying Collectors}
We observe two pathological modes of concurrent copying collectors when challenged with workloads that have high allocation rates but low survival rates.
In such scenarios, the concurrent copying collectors often fall back to STW collections and/or stall the mutators to keep up with the allocation, resulting in poor performance.

One stark result is for xalan, where \Shenandoah has an enormous time LBO of 30.2, about ten times that of \Serial/\Parallel/\GOne.
\ZGC simply failed to run xalan with OOM errors.
Despite the high time LBO, \Shenandoah has a modest cycle LBO of 1.74, which is close to the 1.63 LBO of \Parallel and even slightly better than the 1.78 LBO of \GOne.
To understand this behavior, recall that \Shenandoah and \ZGC rely on tracing to establish liveness, and strictly rely on evacuation for reclamation~(see~\cref{sec:hotspot_gc}).
The consequence is a substantial delay between when an object becomes unreachable and when the unreachable object is reclaimed.
The delay's impact is amplified by the high allocation rates and low survival rates of benchmarks such as xalan and lusearch;\footnote{Both benchmarks allocate multiple GBs per second, and their minimum heap sizes are merely \SI{8}{\mega\byte} and \SI{21}{\mega\byte}, respectively.} because allocation would fail if reclamation did not keep up with the allocation rate, the collector resorts to STW collections.
In this case, it would be more beneficial to use a STW collector in the first place and thus avoid the concurrency overhead~(as we discuss in \cref{sec:misinterpretation}).

By examining the logs from \Shenandoah, we observe two pathological modes.
First, the untimeliness of reclamation causes allocation failures, and \Shenandoah requires STW collection to finish an in-flight concurrent collection~(known as degenerated GCs in \Shenandoah).
Second, in order to avoid STW collections, \Shenandoah throttles allocations by stalling the mutator at allocation sites~(known as pacing in \Shenandoah, or ``allocation stall'' in \ZGC).
Since sleeping threads do not contribute to the cycles consumed, but increase the wall-clock time needed to run a workload, this explains the much higher time cost but modest cycle cost.

\subsection{Misinterpretation of Evaluation Results}
\label{sec:misinterpretation}

Prior work~\cite{methodology-cacm-2008,mmtk-sigmetrics-2004} points out common pitfalls and respective mitigations when evaluating GCs.
We have applied these suggestions where possible~(see \cref{sec:methodology}).
For example, we control the heap size relative to the minimum heap size required to run a given workload.
We reaffirm that the fundamental time--space tradeoff is still applicable in a modern, diverse benchmark suite~(Tables~\ref{table:time_lbo} and \ref{table:cycles_lbo}).
However, these suggestions are not sufficient to avoid misinterpretation of evaluation results, especially in light of modern hardware, concurrent GCs, and emerging latency-sensitive workloads.
In this section, we highlight three important types of misinterpretation.
In the next section, we make recommendations on mitigations.

In this section, we present the misinterpretation by comparing the absolute costs of different GCs shown earlier.
We would like to emphasize that this misinterpretation is also applicable in classic comparative analysis, and using more metrics still improves the evaluations in that context.

\paragraph{Opportunity cost}
Comparing \cref{table:time_lbo} and \cref{table:cycles_lbo}, we notice that collectors have higher cycle LBO than time LBO. %
The only exceptions are \Serial, which is a single-threaded GC, and \Shenandoah at small heap sizes due to pacing, which we discussed previously.
In particular, the cycle LBOs of \Parallel and \GOne are about 0.3$\,\times\,$MDC larger than the respective time LBO at a small \hfac{1.4} heap.
In the extreme case, such as for batik at \hfac{3.0} heap, \Parallel has a mere 1.09 time LBO but a significant 1.99 cycles LBO, the highest among all collectors studied.
This difference reveals that substantial cycle costs can go unnoticed when only the wall-clock time is reported.

Historically, GC performance has been mostly measured using wall-clock time only.
This methodology implicitly assumes that on machines with free cores available, the free cores may be used by GC with no repercussions.
On modern, massively parallel hardware, this assumption can mean that a significant portion of the hardware is dedicated to the GC.
This assumption, however, ignores the incurred opportunity cost.

On multi-tenant hosts, which are increasingly common to increase utilization in datacenters, more CPU cores taken by GC threads mean fewer cores available for other applications on the same host.
That is, when heavily relying on parallelism, a collector will not only run slower when deployed in multi-tenant settings, because fewer cores are available to run GC, but also negatively impact other applications on the same host.
Even when the server has only one application, using fewer cycles to run the application can free up CPU cores and reduce energy consumption.

\begin{table}[tbp]
    \centering
    \caption{Percent of wall-clock time spent in STW pauses averaged over 16 benchmarks.
    Lower is better.
    The best value for each heap size is shown in green.
    Where a collector cannot run all benchmarks at a particular heap size, the corresponding entry is left blank.
    \ZGC consistently outperforms other collectors where it runs.
    }
    \sffamily
\setlength{\tabcolsep}{5.2pt} %
\begin{tabular}{l c@{\hspace{0.5em}} ^r ^r ^r ^r ^r ^r ^r ^r}
\toprule
\textbf{GC} && \hfac{1.4} & \hfac{1.9} & \hfac{2.4} & \hfac{3.0} & \hfac{3.7} & \hfac{4.4} & \hfac{5.2} & \hfac{6.0}\\
\cmidrule{1-1}\cmidrule{3-10}
Ser. && 9.0 & 4.7 & 3.7 & 3.1 & 2.7 & 2.5 & 2.4 & 2.3 \\
Par. && 9.0 & 2.9 & 2.1 & 1.7 & 1.3 & 1.1 & 1.0 & 0.9 \\
G1 && \color{ForestGreen}5.2 & 2.8 & 1.6 & 1.3 & 0.9 & 0.9 & 0.7 & 0.7 \\
Shen. &&  & \color{ForestGreen}0.3 & 0.2 & 0.2 & 0.1 & 0.1 & 0.1 & 0.1 \\
ZGC &&  &  & \color{ForestGreen}0.1 & \color{ForestGreen}0.0 & \color{ForestGreen}0.0 & \color{ForestGreen}0.0 & \color{ForestGreen}0.0 & \color{ForestGreen}0.0 \\
\bottomrule
\end{tabular}

    \label{table:time_stw_portion}
\bigskip
    \centering
    \caption{Percent of total cycles spent in STW pauses averaged over 16 benchmarks.
    Lower is better.
    The best value for each heap size is shown in green.
    Where a collector cannot run all benchmarks at a particular heap size, the corresponding entry is left blank.
    \ZGC consistently outperforms other collectors where it runs.
    }
    \sffamily
\setlength{\tabcolsep}{5.2pt} %
\begin{tabular}{l c@{\hspace{0.5em}} ^r ^r ^r ^r ^r ^r ^r ^r}
\toprule
\textbf{GC} && \hfac{1.4} & \hfac{1.9} & \hfac{2.4} & \hfac{3.0} & \hfac{3.7} & \hfac{4.4} & \hfac{5.2} & \hfac{6.0}\\
\cmidrule{1-1}\cmidrule{3-10}
Ser. &&  \color{ForestGreen}4.3 & 2.1 & 1.6 & 1.4 & 1.2 & 1.1 & 1.0 & 1.0 \\
Par. && 13.1 & 5.3 & 4.0 & 3.4 & 2.8 & 2.6 & 2.4 & 2.2 \\
G1 && 8.1 & 5.0 & 3.3 & 2.8 & 2.1 & 2.1 & 1.9 & 1.8 \\
Shen. &&  &  \color{ForestGreen}0.1 & 0.1 & 0.1 & 0.1 & 0.1 & 0.1 & 0.0 \\
ZGC &&  &  &  \color{ForestGreen}0.0 &  \color{ForestGreen}0.0 &  \color{ForestGreen}0.0 &  \color{ForestGreen}0.0 &  \color{ForestGreen}0.0 &  \color{ForestGreen}0.0 \\
\bottomrule
\end{tabular}

    \label{table:cycles_stw_portion}
\end{table}
\paragraph{Concurrency overhead}

A commonly used metric to tune GC for throughput is the fraction of time spent in GC~(such as the \icode{-XX:GCTimeRatio} flag suggested in the GC tuning guide from the vendor~\cite{tuning-oracle-2021}).
Tables~\ref{table:time_stw_portion} and \ref{table:cycles_stw_portion} show the fraction of time and cycles spent in STW pauses for different GCs.
Similar to Tables~\ref{table:time_lbo} and \ref{table:cycles_lbo}, the results are grouped by heap sizes, and show the geometric means over 16 benchmarks.
Compared with Tables~\ref{table:time_lbo} and \ref{table:cycles_lbo}, we can see that the classic methodology of estimating the GC costs from the time/cycles spent in GC pauses is highly problematic, especially for the concurrent collectors.
In particular, the two concurrent copying collectors spend a negligible fraction of time/cycles in GC pauses, but have enormous LBOs.%

With fixed hardware resources, concurrent GC threads compete with mutator threads, e.g., for cache capacity and memory bandwidth, resulting in more severe effects for more parallel workloads.
In other words, concurrent GC costs are high not only from the expensive mechanisms they use, such as read and write barriers, but also from resource contention.

\paragraph{Low pause $\ne$ low latency}
A commonly used metric to tune GC for latency-sensitive applications is the maximum GC pause time~(such as the \icode{-XX:MaxGCPauseMillis} suggested in the GC tuning guide from the vendor~\cite{tuning-oracle-2021}).
It is well known that the pause time is a poor metric to assess GCs for latency-sensitive applications~\cite{mmu-pldi-2001}.
We reaffirm this here, and highlight the importance of using more metrics.
The following example shows that fixating on limited metrics~(\eg, pause times) can even make a language implementer unintentionally work against the motivating goal~(the responsiveness of the application).

\begin{figure}[tbp]
    \centering
    \includegraphics[width=\linewidth]{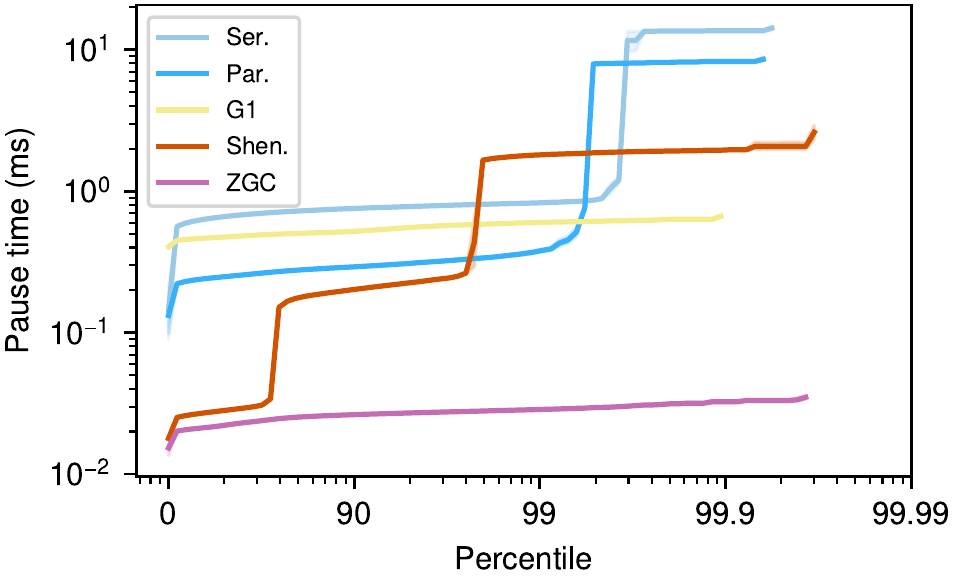}
    \caption{GC pause time for lusearch in a \hfac{3.0} heap.
    Each line and its shade show the mean and 95\,\% CI over 20 invocations.}
    \label{fig:lusearch_pause}
\end{figure}
\begin{figure}[tbp]
    \setlength{\belowcaptionskip}{-1.5ex}
    \centering
    \includegraphics[width=\linewidth]{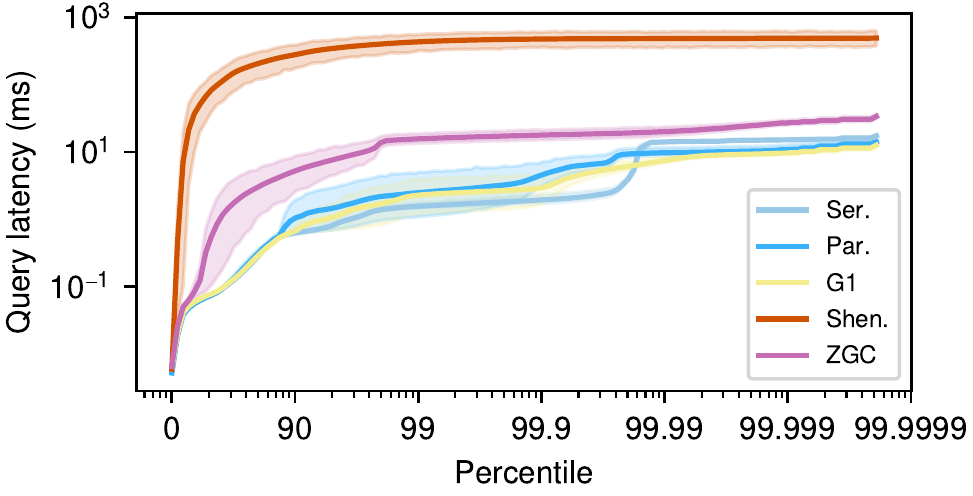}
    \caption{Metered latency for lusearch in a \hfac{3.0} heap.
    Each line and its shade show the mean and 95\,\% CI over 20 invocations.}
    \label{fig:lusearch_latency}
\end{figure}

Figures~\ref{fig:lusearch_pause} and \ref{fig:lusearch_latency} show the distribution of pause times and query latencies~(using metered latency; see \cref{sec:methodology}) of different GCs running the lusearch benchmark at \hfac{3.0} heap.
The low-pause collectors~(\Shenandoah and \ZGC) indeed achieve better pause times in general, with \ZGC consistently having the lowest pause time for all percentiles, while \Shenandoah has lower pause times than the other three GCs under the 90th percentile.
However, low pause times do not automatically confer low application latencies.
Indeed, both \Shenandoah and \ZGC have worse (by factors of 10--100$\times$) query latencies than the other three collectors.

Application latencies are affected by GC pauses---both by their durations and frequency.
Short but frequent pauses will not impact the distribution of pause times, but can certainly impact application latency.

Importantly, application latencies are also functions of mutator performance.
As discussed in previous sections, concurrent copying GC can affect mutator performance through expensive mechanisms, such as read barriers, and through resource contention.
In the case of the pathological modes we discussed, \Shenandoah and \ZGC throttle mutator threads when the GC cannot keep up with allocation.
Such throttling indeed avoids triggering a STW collection, keeping each GC pause short.
However, it comes at great cost: if the mutators are sufficiently throttled, both the application latency and throughput will be worse than a simple STW GC, as evidenced in the above graphs and \cref{table:time_lbo_3023}.

\subsection{Recommendations}
Drawing on these  observations, we make two recommendations for improving GC evaluation in future research.
First, the distillation methodology should be used to report the costs of GCs.
This helps us better understand the scale of the impact of GC on the program execution.
Second, a richer set of performance and cost metrics should be used when evaluating GCs.
At a minimum, both the wall-clock time and the CPU cycles used should be reported.
Any additional metric can help us understand the performance characteristics of different GCs better.
This includes measuring application latency for applications with latency requirements, instead of using pause times as a surrogate.

\section{Conclusion}
In this paper, we identify two important problems in empirical evaluation of GCs: unclear costs, and easy-to-misinterpret results presented using limited metrics.
To address these problems, we first devise the distillation methodology to place an empirical lower bound on the costs of GCs for any given cost metric.
Then, we use the distillation methodology to measure the costs of five production collectors in OpenJDK 17.
We find that these GCs incur substantial costs: (as an underestimate) at least
7--82\,\% more time and 6--92\,\% more cycle overhead relative to the intrinsic application cost for a modest \hfac{2.4} heap.

Our results also show how a lack of diverse cost/performance metrics can lead to misinterpretation of GC evaluations, hindering GC development.
We identify three important types of misinterpretation: neglecting opportunity cost, neglecting concurrency overhead, and using GC pause times as a proxy metric for application latency.
These types of misinterpretation can be mitigated by including more metrics in the evaluation.

Our findings reveal substantial costs in production GCs, highlighting opportunities for future research on low-latency collectors with low costs.
We recommend using more metrics when evaluating GCs, because this helps reveal tradeoffs and limitations of GCs that often go unnoticed.
The distillation methodology is language and runtime agnostic, and thus can benefit GC research on a wide range of managed languages and platforms, such as Go, \ .NET~(for C\# and other CLI languages) and V8~(for JavaScript).

\bibliographystyle{IEEEtran}
\bibliography{paper.bib}
\balance

\end{document}